\documentclass[journal,twoside,web]{ieeecolor}
\usepackage{tmi}

\usepackage{amsmath,amssymb,amsfonts}
\usepackage{graphicx}
\usepackage{textcomp}
\usepackage{footnote}
\def\BibTeX{{\rm B\kern-.05em{\sc i\kern-.025em b}\kern-.08em
    T\kern-.1667em\lower.7ex\hbox{E}\kern-.125emX}}
    
\usepackage{bbm}
\usepackage{ctable}
\usepackage{multirow}
\usepackage{subcaption}
\usepackage[normalem]{ulem}

\usepackage{color, colortbl}
\definecolor{LightCyan}{rgb}{0.88,1,1}

\newcolumntype{g}{>{\columncolor{LightCyan}}c}

\graphicspath{{figures/pdf/}{figures/png/}}
\DeclareGraphicsExtensions{.pdf,.png}

\newcommand{\etal}{\mbox{\emph{et al.}}}
\newcommand{\ie}{\mbox{\emph{i.e.,}}}

\usepackage{tabu}
\usepackage{comment}
\usepackage{cite}
\usepackage{hyperref}  

\begin{document}
\title{Learning With Context Feedback Loop for Robust Medical Image Segmentation}
\author{Kibrom~Berihu~Girum, Member, IEEE, Gilles~Cr\'ehange, and~Alain~Lalande \vspace{-0.5cm}
\thanks{Manuscript received December 4, 2020; revised January 19, 2021; accepted February 16, 2021. (Corresponding author: Kibrom Berihu Girum.)}
\thanks{K.B. Girum is with the Imaging and Artificial Vision (ImViA) Research Laboratory, University of Burgundy, 21000 Dijon, France, and also with the Department of Radiation Oncology, Centre Georges François Leclerc (CGFL), 21000 Dijon, France. (Corresponding author's email: kibrom2b[at]gmail.com.)}
\thanks{G. Cr\'ehange is with the Department of Radiation Oncology, Institute of Curie, 75005 Paris, France, also with Imaging and Artificial Vision (ImViA) Research Laboratory, University of Burgundy, 21000 Dijon, France, and also with the Department of Radiation Oncology,
 Centre Georges François Leclerc (CGFL), 21000 Dijon, France.(Email: gilles.crehange[at]curie.fr)}
\thanks{A. Lalande is with the Department of Medical Imaging, University Hospital of Dijon, 2100 Dijon, France, and  also with Imaging and Artificial Vision (ImViA) Research Laboratory, University of Burgundy, 21000 Dijon, France. (Email: alain.Lalande[at]u-bourgogne.fr)}
\thanks{Copyright \textcopyright 2021 IEEE. Personal use is permitted. For any other purposes, permission must be obtained from the IEEE by emailing pubs-permissions@ieee.org.}
\thanks{This is the author's version of an article that has been published in this journal. Changes were made to this version by the publisher prior to publication. The final version of record is available at \\ \url{https://doi.org/10.1109/TMI.2021.3060497}}
}

\maketitle

\begin{abstract}
 \label{sec:abstract}
Deep learning has successfully been leveraged for medical image segmentation. It employs convolutional neural networks (CNN) to learn distinctive image features from a defined pixel-wise objective function. However, this approach can lead to less output pixel interdependence producing incomplete and unrealistic segmentation results. In this paper, we present a fully automatic deep learning method for robust medical image segmentation by formulating the segmentation problem as a recurrent framework using two systems. The first one is a forward system of an encoder-decoder CNN that predicts the segmentation result from the input image. The predicted probabilistic output of the forward system is then encoded by a fully convolutional network (FCN)-based context feedback system. The encoded feature space of the FCN is then integrated back into the forward system's feed-forward learning process. Using the FCN-based context feedback loop allows the forward system to learn and extract more high-level image features and fix previous mistakes, thereby improving prediction accuracy over time. Experimental results, performed on four different clinical datasets, demonstrate our method's potential application for single and multi-structure medical image segmentation by outperforming the state of the art methods. With the feedback loop, deep learning methods can now produce results that are both anatomically plausible and robust to low contrast images. Therefore, formulating image segmentation as a recurrent framework of two interconnected networks via context feedback loop can be a potential method for robust and efficient medical image analysis.  
\end{abstract}

\begin{IEEEkeywords}
CNN, Feedback loop, MRI, Ultrasound, CT.
\end{IEEEkeywords}

\section{Introduction}
\label{sec:introduction}
\subsection{Motivation and background}
\IEEEPARstart{M}{edical} image segmentation is often profoundly important in clinical image analysis and image-guided interventions. It involves partitioning a medical image into multiple areas, where these areas can be used for better clinical analysis or clinical target visualization. For example, in prostate radiotherapy, accurate clinical target volume segmentation from magnetic resonance (MR), ultrasound (US), or computed tomography (CT) images is often essential in computer-aided diagnosis, therapy, and post-therapy analysis of prostate cancer \cite{davis2012american} \cite{girum2018inferring}. Indeed, it is critical to select patients for a specific treatment, guide source delivery during the intervention, and compute the dose distribution using MR, US, and CT images, respectively \cite{davis2012american}. Similarly, in cardiac image analysis, accurate segmentation of the heart structures such as the left and right ventricular cavities, and the myocardium is essential to calculate the volume of the cavities at end-diastolic and end-systolic phases and the left ventricular myocardial mass \cite{petitjean2011review} \cite{bernard2018deep}. Inner ear structures segmentation such as the cochlea, vestibule, and semi-circular canals from preoperative CT images can be used in the treatment of patients with hearing impairments. In cochlear implant surgery, preoperative cochlea segmentation from CT images is useful for determining the length of the implanted electrode and improving the insertion guidance \cite{gerber2017multiscale}. Other important medical image analysis applications using image segmentation includes in 2D echocardiography \cite{jafari2019echocardiography} and brain image segmentation \cite{cciccek20163d} among others.

Despite the necessity of accurate and robust image segmentation in clinical routines, it is often challenging. Indeed, it highly depends on the imaging modality and the target (such as anatomy or tumor area). The main challenges in developing accurate and automatic medical image segmentation methods include a wide range of patient characteristics, low-contrast inherent imaging characteristics of some imaging modalities, artifacts (e.g., metallic artifacts in CT), respiratory motion, limited data in medical image analysis, and significant variation of shape and size of organs among cases \cite{ghose2012survey} \cite{girum2020fast}. Unfortunately, manual segmentation is frequently prone to subjective errors and is time-consuming. It might not always be easy to analyze multiple sequences of examinations manually. Researchers have applied different methods to address these challenges, thereby better analyzing the medical images and improving case outcomes. For example, several groups have approached medical image segmentation using contour and shape detection, deformable models, conventional machine learning, and deep learning approaches \cite{ghose2012survey} \cite{litjens2017survey}. 

In the early times, conventional medical image segmentation approaches are often based on edge detection and fitting predefined parametric shapes, level-set methods, and shape models \cite{petitjean2011review} \cite{ghose2012survey}. 
Many classical machine learning-driven methods (supervised and unsupervised) are also proposed to tackle this challenging problem. Although traditional machine learning approaches like active shape, atlas methods, and statistically supervised and unsupervised methods are still prevalent \cite{litjens2017survey}, they often involve careful integration of hand-crafted image features. These hand-crafted features need to represent the input image. However, designing or engineering distinctive image features by human experts can be difficult and sometimes impossible. Moreover, manually designed distinctive features on a given task might not be easy to adapt to new cases, which is the main problem in developing a general method that can be used to extract the distinctive image features \cite{ghose2012survey} \cite{litjens2017survey}.

In this regard, deep learning methods based on convolutional neural networks (CNN) have emerged as an alternative and reliable solution \cite{litjens2017survey}. These CNN-based methods can automatically learn to extract the hierarchical distinctive image features. It avoids the need to develop hand-crafted features besides the generic characteristics of these extracted features. In fact, CNN-based approaches have achieved the state of the art (SOTA) performance in various tasks such image classification \cite{krizhevsky2012imagenet}, object detection \cite{he2017mask}, segmentation \cite{ronneberger2015u} \cite{long2015fully}, registration \cite{xu2019deepatlas} \cite{li2019hybrid}, and other tasks \cite{goodfellow2014generative} \cite{simonyan2014very}. For example, fully convolutional neural networks are trained on pixel-to-pixel transformations to extract high-level image features and predict the outputs from any arbitrary size inputs \cite{long2015fully}. Following the success of these approaches in image and video processing, Simonyan \etal~\cite{simonyan2014very} proposed to increase the convolutional network depth via a network called VGG16. In this approach, convolutional layers are stacked one after the other to extract more distinctive image features.

Motivated by the promising results of the VGG16 network (deep encoder like structure), Badrinarayanan \etal~\cite{badrinarayanan2017segnet}, proposed to expand it into a deep fully convolutional neural network (FCN) architecture for semantic pixel-wise segmentation. It employs a trainable encoder network for high-level feature extraction and a corresponding decoder network. The decoder network maps the low-resolution encoder feature maps to a full input resolution feature maps for pixel-wise classification using up-sampling. This up-sampling operation is then modified to be a learning-based deconvolution network in the work of Noh \etal~\cite{noh2015learning}. Szegedy \etal~\cite{szegedy2015going} also proposed a new design of CNN architecture to increase the depth and the width of CNN-based methods. As CNN architectures go deeper (\ie~large number of convolutional layers are stacked one after the other), residual connections are inherently crucial for the training \cite{he2016deep} \cite{szegedy2017inception}. It smooths training and avoids loss of high-level features during sampling or pooling operations.  

Ronneberger \etal~\cite{ronneberger2015u} modified the FCN proposed by Long \etal~\cite{long2015fully} to propagate contextual information from the encoder 
into the decoder. 
It is done by connecting the encoder with the decoder through skip connections 
that creates a U-shaped architecture (and named U-Net). 
Similarly to the residual networks \cite{he2016deep}, this skip-connection recovers spatial information (such as localization information) that could be lost during the consecutive striding and convolutional operations. Several groups have modified the U-Net architecture for various medical image analysis applications \cite{litjens2017survey}. \c Ciçek \etal~\cite{cciccek20163d} adopted the 2D U-Net for volumetric (3D) segmentation. Tu \etal~\cite{tu2009auto} introduced auto-context U-Net in which they applied parallel 2D convolutional layers for the axial, coronal, and sagittal planes of brain images. 

Most previous works based on encoder-decoder (or U-Net) architectures have been focused on improving encoder's feature extraction ability either by constraining the segmentation results \cite{oktay2017anatomically}, incorporating the prior knowledge such as shape models \cite{girum2019deep} \cite{zotti2018convolutional}, or using transfer learning from pre-trained networks \cite{raghu2019transfusion}  \cite{gu2019net}.  

Meanwhile, there has been an increased interest of researchers to incorporate the shape of anatomical structures into U-Net like architectures using multi-task learning \cite{zotti2018convolutional} \cite{zeng2019liver} \cite{girum2020deep}. These approaches aimed at solving the common limitations of encoder-decoder architectures in capturing the structural information and interdependence of the output while training from pixel-wise objective functions \cite{litjens2017survey}. Consequently, several groups approached the incorporation of shape prior in image segmentation using statistical shape modeling 
\cite{zotti2018convolutional} or learning-based modeling \cite{oktay2017anatomically}  \cite{zeng2019liver}
\cite{chen2019learning} \cite{ravishankar2017learning}.  These approaches generally showed a relative improvement over the methods without the shape prior. However, these approaches often require explicit prior knowledge of the target. It needs the modeling of the prior knowledge, for example, anatomical shapes, and then embed it into the U-Net architectures to constrain the learning process \cite{jafari2019echocardiography} \cite{oktay2017anatomically} \cite{zotti2018convolutional}. Other approaches include post-processing methods based on either denoising \cite{larrazabal2020post} or variational auto-encoders \cite{painchaud2020cardiac} and showed an improvement in the plausibility of the results. However, they are not free of limitations. Post-processing methods do not see the original input image. Thus, they might not always produce accurate results from a given erroneous segmentation \cite{painchaud2020cardiac}.

In encoder-decoder based deep learning methods, although the encoder is essential to extract high-level and distinctive input image features, it could be challenging due to the often inconsistent contrasts and potential artifacts associated with the inherent limitations of medical imaging devices. Thus, similar to the residual \cite{he2016deep} and the long skip connection \cite{ronneberger2015u} approaches, different alternative methods have been proposed to improve the contextual information propagation from the encoder to the decoder. These methods use either gated attention-based networks \cite{anderson2018bottom} or recurrent neural networks \cite{schuster1997bidirectional}. They can be used either to highlight salient image features by capturing richer contextual dependencies \cite{schlemper2019attention} \cite{sinha2020multi} or to recurrently recover information loss from the consecutive sampling and re-sampling operations \cite{li2018referring} \cite{chen2016combining} \cite{alom2019recurrent} \cite{wang2019recurrent} in the feed-forward learning approach. Recurrent based networks use a similar concept to residual networks. Firstly, they are used to accumulate features that lead to better feature representation. Secondly, they are used to smooth the training \cite{alom2019recurrent}. Other approaches proposed to apply additional paths to the encoder-decoder architecture \cite{lin2017refinenet} \cite{wangfrnet}. Most of these approaches are based on CNNs in the feed-forward training approach and could be biased towards recognizing the texture features than the contextual or shape features \cite{geirhos2018imagenet}.

Nonetheless, capturing the contextual information, the inter-region relationship, and implicitly learning the prior knowledge of a target remained a wide interest of researchers in deep learning-based medical image analysis. Still, encoder-decoder based image segmentation methods do not get the second chance to look at the segmentation results, except through the gradient backpropagation from a defined pixel-wise objective function. Indeed, training from only pixel-wise objective functions can lead to the loss of spatial information, less output pixels interdependence, and less inter-region relationship representation and hence sometimes, produce unrealistic segmentation results \cite{oktay2017anatomically}. A recent study has also demonstrated that CNNs are strongly biased towards recognizing textures rather than shapes \cite{geirhos2018imagenet}. Thus, capturing the output pixels' interdependence would enable the network to learn distinctive image features. The network can learn to extract texture and contextual similarity between the same labeled pixels and the difference between differently labeled neighboring pixels, thereby producing realistic segmentation.  

In this regard, the decoder should be intelligent enough to improve the classification of each pixel. To do so, the decoder needs to capture the contextual information, inter-region relationship, and distinguish errors that could be introduced during the consecutive striding, convolution, and up-sampling operations. Indeed, feeding back the predicted probabilistic output could help the decoder extrapolate contextual information and correct mistakes over time, thereby improving prediction accuracy. The idea of a feedback loop is indeed used in the control system's theory, in which the system's output error signal is used to make adjustments on the input signal. Alternative studies on generative adversarial neural networks (GANs) have also recently gained an interest in using a feedback loop to improve the model's learning capacity \cite{shama2019adversarial} \cite{huh2019feedback}. 

Therefore, in this study, to address the difficulties of CNNs in capturing the contextual information and the inter-region relationship, and to implicitly integrate the prior knowledge into the feed-forward learning process, we formulated image segmentation as a recurrent framework using two systems (named Learning with context FeedBack system (LFB-Net)). One can notice that we used the term system to refer to the two different interconnected networks. The forward system, a modified U-Net architecture, learns and predicts a probabilistic output from the raw input image. The probabilistic refers to the pixel levels' softmax assignment at the output of the network. The second system, a fully convolutional network (FCN), named hereafter feedback system, transforms the predicted probabilistic output of the forward system into another spatial higher-level representation. The transformed high-level representation is then integrated back into the feed-forward learning path of the forward system. This feedback strategy, as we will show, would allow the main segmentation module to improve the performance over time. More importantly, the feedback system mitigates the conditions in which the forward system can potentially fail. 
\subsection{Contributions}
In this study, we present the first deep learning formalism using a feedback loop in encoder-decoder based image segmentation. 
To this end, we designed the image segmentation problem as a recurrent framework. The image segmentation process of a network is conditioned by its previous segmentation results' latent space and a spatial response map of another network that enables it to improve the performance over time.

Our main contributions can be summarized as follows: 
\begin{enumerate}
\item  We introduce a feedback looping system to enhance the feed-forward learning process of encoder-decoder CNNs for medical image segmentation. 
\item We model the image segmentation problem as a two systems task, from system one to system two approach. We integrated the proposed feedback looping system with a modified U-Net architecture (in the forward system) as a regularizer that enables the network to attend and fix segmentation uncertainties over time.

\item The proposed method, LFB-Net, is validated on different medical image segmentation applications. Specifically, we evaluated it on short-axis cardiac-MRI, long-axis echocardiographic images, CT of the prostate, and CT of the inner ear image segmentation applications. Our extensive experiments and ablation studies on clinical databases support that our method consistently outperforms the SOTA methods at a reduced computational cost. As will be shown in the result section, it yielded accurate and plausible results without the need to incorporate shape prior or apply post-processing methods.
\end{enumerate}

The rest of this paper is organized as follows. Section II introduces the proposed method (LFB-Net). In section III, we present the experimental results and discussions. Finally, we draw our conclusions in section IV.  

\section{Methodology}
The main objective of a supervised FCN network is to learn how to best predict the target output $y$ from a given input image $x$, \ie~mapping of the input image into the labeled target ($f: x \mapsto y$). The network $f$ thus learns through a back-propagation algorithm from a defined loss function $L(y, \hat{y})$ (which is often an error between the model's predicted $\hat{y}$ and the reference $y$ values). Thus, it learns how to compute the model's neural network weights $w$ and thereby best map the input image into the output target image, \ie~$\hat{y} = f(x;w)$. 

\begin{figure*}[!t]
	\centering
	\includegraphics[width=1\linewidth, height=0.5\linewidth]{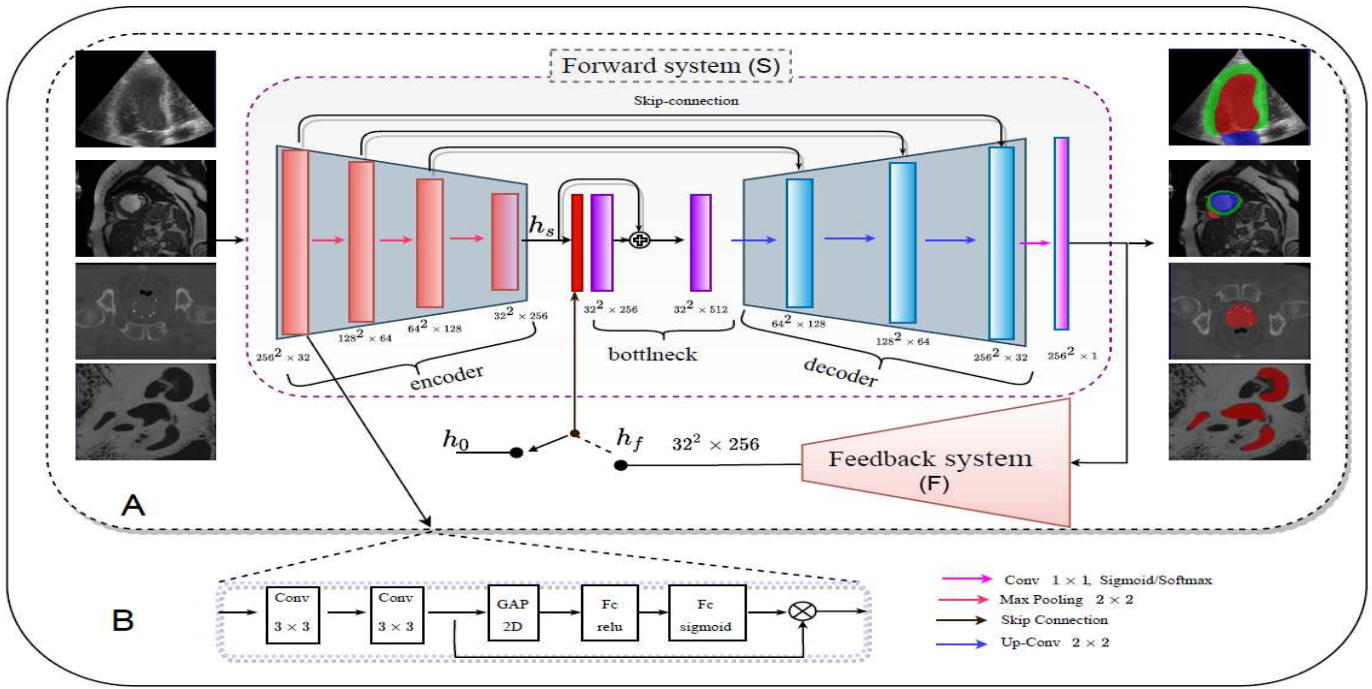}
	\caption{Framework of the proposed method: (A) the architecture of LFB-Net. In which the encoder, decoder, and bottleneck are the component of the forward system (U-Net architecture), and (B) the convolutional building block of the encoder as well as the decoder. The feedback system's output (\ie~$h_{f} =32^{2} \times 256$) is concatenated with forward system encoder's output (\ie~$h_{s}$). The switch indicates the alternative training strategy of the two systems. From top to bottom: Echocardiography long-axis view, cine-MRI short-axis view, prostate CT, and CT of the inner ear in the left and segmentation results on the right.}
	\label{fig_sim}
\end{figure*}

In our model, LFB-Net consists of two major parts: the main segmentation module (forward system) and the feedback system (Fig. \ref{fig_sim}). Each module is also composed of an encoder and a decoder. An encoder-decoder based FCN architectures learn the distinctive features of the datasets from a defined objective function $L_{f} (y, f_{d}(f_{e}(x)))$ \cite{long2015fully}. Here, the functions $f_{e}$ and $f_{d}$ are defined as the encoder and the decoder mapping functions, respectively (\ie~$f_{e}:x \mapsto h$, and then $f_{d}: h \mapsto y$). The intermediate representation of the input image is then $f_{e} (x) = h$, often named latent space or higher feature space. It is another transformed spatial representation of the input image $x$. We have used this representation to describe our method in the following sections. 
\subsection{Forward system}
\label{sec:segmentationmodule}
We formulate the main segmentation network $S$ (forward system in Fig. \ref{fig_sim}) as two main parts: a segmentation encoder $S_{e}$
which encodes the raw input image into high feature space, and a decoder $S_{d}$ which then decodes the encoded features into the target labels. Specifically, the encoder network $S_{e}$  transforms the input image $x$ into the encoded features space $h$, \ie~$S_{e} : x \mapsto h$, and the decoder transforms the intermediate representation $h$ into the desired label $y$, \ie~$S_{d} : h \mapsto y$. This consecutive stage is equivalent to 
\begin{equation}
\label{eq:segmentation}
\hat{y} = S(x) = S_{d} (S_{e} (x)),
\end{equation}
where $x \in \mathbb{R}^{W\times H} $ is the input image and $\hat{y} \in \mathbb{R}^{W\times H}$ is the predicted output label. Here, $W$ and $H$ indicate the original image width and height respectively. Thus, the latent space dimension of this network is $ S_{e}(x)=h_{s} \in \mathbb{R} ^{W/d \times H/d \times C}$, where $C$ indicates the number of feature channels at the latent space while $d$ is the depth of the network. 

In order to not lose spatial information during down-sampling, this network is composed of skip connections. It is typically similar to the U-Net architecture \cite{ronneberger2015u}, except the encoder and decoder are designed so that they can be trained jointly and separately, as will be shown in section \ref{sec:segmentationwithfeedbacksystem}.

\subsection{Feedback system}

The feedback system aims to provide a second chance for the forward system's decoder network to look back on its predicted output. This approach could be made by recurrently formulating the network and merging their full-resolution outputs at the last layer of the decoder. However, providing a more high-level representation of the predicated output with small correction using the feedback system (\ie~spatial feedback) would enhance the decoder's learning capacity \cite{huh2019feedback}. Thus, the forward system's probabilistic output is fed to a new FCN ($F$) architecture (the feedback system), which transforms the predicted probabilistic output into another high-level feature space. This approach of correcting and feeding back the output can be considered as an analogy to the teacher correcting the exam of a student and then providing back the correct answers. Therefore, this strategy allows the forward system to attend to specific regions of its previous results and improve prediction accuracy. 

Lets divide the feedback system $F$ into two: the encoder $F_{e}$ which encodes the forward system's probabilistic output $\hat{y}$ into a high-feature space $h_{f}$ ( \ie~$F_{e} : \hat{y} \mapsto h_{f}$) and the decoder $F_{d}$ then transforms the high-feature space $h_{f}$ into desired label $y$ (\ie~$F_{d} : h_{f} \mapsto \hat{\hat{y}} $).  As in (\ref{eq:segmentation}), this process can be written as:

\begin{equation}
\hat{\hat{y}} = F(\hat{y}) = F_{d} (F_{e} (\hat{y})),
\label{eq:feedbacksystem}
\end{equation}

where $\hat{\hat{y}}$ is the predicted probabilistic output label from the feedback system, and $\hat{y}$ is the predicted probabilistic output label from the forward system. 

\subsection{Integration}
\label{sec:segmentationwithfeedbacksystem}
We need to integrate the feedback system $F$ with the forward system $S$. Thus, we designed the integration of the two systems as a recurrent process. This recurrent process is formulated as follows. Firstly, the segmentation module produces a label map $\hat{y_{i}}$ (at iteration $i$) using (\ref{eq:segmentation}). Secondly, the feedback system predicts a label map $\hat{\hat{y_{i}}}$ taking $\hat{y_{i}}$ as input from the output of the forward system. Now, we can write the feedback system's encoder response map $F_{e}$ as:

\begin{equation}
\label{eq:feedbackintegration}
h_{f_{i}} = F_{e} (\hat{y_{i}})
\end{equation}

where $h_{f_{i}} \in \mathbb{R} ^{W/d \times H/d \times C}$ is high-feature space representation or latent space of the feedback system at iteration $i$. It has the same size as the forward system's encoder output, $h_{s_{i}} \in \mathbb{R} ^{W/d \times H/d \times C}$. The decoder ($F_{d}$) then transforms $h_{f_{i}} $ into the output label map $\hat{\hat{y_{i}}}$. 

Thirdly, we use the high-feature space of feedback information $h_{f_{i}}$ along the encoded input from forward system $h_{s_{i}}$ as input to the decoder $S_{d}$, at iteration $i+1$.  Therefore, now we can redefine (\ref{eq:segmentation}) at $i+1$ as:
\begin{equation}
\hat{y}_{i+1} = S_{d_{i+1}} \big ( h_{s_{i}},  F_{e_{i}} (\hat{y}_{i}) \big) 
\end{equation} 

which can be written as
\begin{equation}
\hat{y}_{i+1} = S_{d_{i+1}} \big ( h_{s_{i}}, h_{f_{i}} \big) \
\label{eq:decoder_feedbavck}
\end{equation} 
As our system aims to incorporate the context spatial feedback in the feed-forward learning process, at the first stage we set the feedback latent space $h_{f_{0}} = 0$ (\ie~$h_{f_{0}}= h_{0}$). Then in the second stage (\ie~at $i=i+1$), we replace $h_{f_{0}}$ by the feedback latent space $h_{f_{i}}$. This is indicated in Fig. \ref{fig_sim} as switch between $h_{0}$ and $h_{f}$. 

\subsection{Training strategy}
\begin{figure}[h]
	\centering
	\includegraphics[width=1\linewidth]{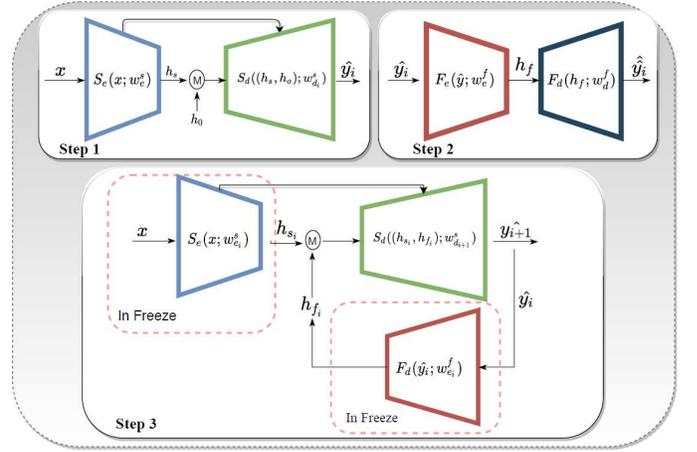}
	\caption{Training scheme of the proposed LFB-Net method. For a given iteration $i$: Step 1: train the forward system;  Step 2: train the feedback FCN system. And for the next iteration $i+1$: Step 3: train the forward system's decoder network with the feedback loop. In step 3 only the neural network weights of $S_{d}(h_{s}, h_{f})$ $( \ie~w_{d_{i+ 1}}^{s})$ are updated. In this case, while the encoder ($S_{e}(x;  w_{e_{i}}^{s})$) is in predicting mode, the feedback loop ($F_{d} (\hat{y}_{i}; w_{f_{e_{i}}})$) is used to regulate the forward system's decoder ($S_{d}$) by feeding back its predicted probabilistic output in step 1. M is the merging block for $h_{f}$ and $h_{s}$. In freeze refers that the network is predicting the output using the trained weights.}
	\label{fig_training} 

\end{figure}

The integration and training strategy of the forward system and the feedback system can be summarized as follows:
\begin{enumerate}[]
	\item Train the neural network weights of the forward system, $w^{s}_{i}$, considering the raw input image $x$ and zero feedback latent space 
	(\ie~$h_{f_{i}}=0$) as inputs, and the ground truth labels $y$ as output. 
	\item Train the neural network weights of the feedback system, $w^{f}_{i}$, considering the input from the predicted output of the forward system's decoder network $\hat{y}$, and the ground truth label $y$ as output as in (\ref{eq:feedbacksystem}). 
	\item Train the neural network weights of the forward system's decoder part only $w^{s}_{d_{i+1}}$, taking the inputs from previously extracted high-level features from the raw input image in step 1 (\ie~$h_{s_{i}}$) and the feedback latent space $h_{f_{i}}$ from the feedback system in step 2 as in (\ref{eq:decoder_feedbavck}). Here, the forward system's and the feedback system's encoder are designed to predict (\ie~freeze) from previously learned and updated weights during step 1 and step 2, respectively.
	\item While not converged, repeat the above steps. The convergence is determined by the change in the validation loss at the output of the forward system. These steps are also shown in Fig. \ref{fig_training}. 
\end{enumerate}

Note that each step is trained at a time until the network sees the whole training dataset. During the training phase, the feedback system provides feedback to the forward system, but the decoder part is discarded during the testing phase.

\subsection{Training Loss function}
To train our model, we used an average of binary cross-entropy and Dice coefficient loss functions as:  
\begin{equation}
\label{equation:sum_of_loss}
L_{total} = \frac{1}{2} \times (L_{1} + L_{2})
\end{equation} 
where $L_{1}$ and $L_{2}$ are the cross-entropy and the Dice loss functions, respectively. The cross-entropy $L_{1}$ is computed as :
\begin{equation}
L_{1}= - \sum_{k = 1}^{c} \sum_{i=1}^{I}\log{\bigg[\frac{e^{s(k, i)}} {\sum _{i} e^{s(k, i)}} \bigg]}
\end{equation}
where $s(k, i)$ is the probabilistic feature maps at pixel $i \in I$, belonging to the pixel class $k \in 1,2,3, ...c$ ($c$ number of classes), and $I$ being the number of pixels in the training batch. The Dice coefficient loss is calculated as:
\begin{equation}
L_{2} =  1 - \frac{1}{\sum_{k}^{} \gamma_{k}} \bigg[ \sum_{k}{\gamma_{k}} \frac{2 \times \sum_{i\in I} u_{i}^{k} v_{i}^{k}}{\sum_{i\in I}^{} u_{i}^{k} + \sum_{i\in I}^{} v_{i}^{k}  } \bigg]
\end{equation}

where $u$ is the predicted output of the network, $v$ is a one-hot encoding of the ground truth segmentation map, $\gamma$ is the weight associated to class $k \in {0,1,2, ...}$, being the pixel class.  Both $u$ and $v$ have shape $I$ with $i \in I$ being the number of pixels in the training batch. The loss function was the same for both the forward and the feedback systems. 

\subsection{Network architecture}
\label{subsec: Networkarchitecture}
The complete network architecture is shown in Fig. \ref{fig_sim}. In the forward system, the encoder's blocks consist of repeated applications of $3\times3$ convolutions, exponential linear unit (ELU) activation, batch normalization, and followed by a squeeze-and-excitation network \cite{hu2018squeeze} (Fig. \ref{fig_sim} (B)). Moreover, we applied a $2\times2$ max pooling operation with stride 2 for downsampling. Contrarily, the decoder is composed of a $2\times2$ up-convolution and concatenation layer, followed by the same block as in the encoder (Fig. \ref{fig_sim} (B)). 

The feedback system is a fully convolutional network architecture with a learning deconvolution network \cite{noh2015learning}. It consists of a repeated application of  $3\times3$ convolutions, ELU activation, batch normalization in both the encoder and decoder blocks. We selected a concatenation layer to merge the latent spaces from the two systems. Please refer to the supplementary material for a detailed description of the architectures and hyper-parameter values used in this work.

\section{Materials and experiments}
\label{sec:experimentalsetupandresults}
In this section, firstly, we introduce implementation, training, and testing configurations. Secondly, we present the experimental studies on four datasets. The datasets are selected to demonstrate our method's performance on various medical image segmentation tasks such as single and multi-structure segmentation. Specifically, prostate volume segmentation from CT images and cardiac multi-structure segmentation from 2D echocardiography and 3D cardiac-MR images are used. Moreover, to demonstrate our method's performance for applications involving irregularly shaped multiple targets per image and more class imbalances, we applied it for inner ear segmentation from $ \mu$CT. Besides, we performed an extensive ablation study to analyze and validate the design of the method. These experiments focus on demonstrating the importance of learning with a feedback loop for accurate and robust medical image segmentation tasks.

\subsection{Experimental setup}
The system was implemented and developed in Python, Keras API, with a Tensorflow backend. The weights of the model were updated by using (\ref{equation:sum_of_loss}),  and an ADAM optimizer with a learning rate of $10^{-3}$ until convergence \cite{kingma2014adam}. All networks, including the baseline models, were trained from scratch with an early stop of 100 and batch-size of 10. 
We mostly divided the dataset into 70\% training and the rest 30\% for testing. For each clinical datasets which involve volumetric computation, the division was done based on the patient cases. The model is trained on the given training dataset and then tested on unseen new image cases. The model's hyper-parameters were optimized only from the validation dataset. 

The proposed network's parameters were the same for all segmentation applications, except for the output activation function and the number of output feature channels. For the single label segmentation (\ie~for prostate and inner ear), it was a sigmoid activation function. For the multi-label segmentation (\ie~for cardiac cine-MRI and ultrasound with four channel outputs), it was a softmax activation function. In all experiments, the intensity values were normalized according to each dataset's mean and standard deviation. It allows the model to learn the optimal model parameters quickly. 

The proposed method was compared with other methods such as U-Net \cite{ronneberger2015u}, residually connected U-Net (ResU-Net) \cite{he2016deep}, FCN \cite{noh2015learning}, Post-DAE \cite{larrazabal2020post}, and attention gated network (AGN) \cite{schlemper2019attention}. We chose these networks as they have a similar architecture to our method. For the public datasets, we evaluated our method on the testing data. We also compared the proposed method with other methods such as \cite{oktay2017anatomically}, \cite{painchaud2020cardiac}, and \cite{leclerc2019deep} on the public data. We extensively experimented during our method's ablation study with and without a feedback loop such as using only the forward system (FS). All settings were then the same for the baseline methods. Indeed, the baseline method's hyper-parameters were optimized to yield the best results. 

We considered both volume and distance metrics to evaluate the methods, such as the Dice similarity coefficient (DSC), Hausdorff distance (HD), and relative volume difference (RVD). For the volumetric datasets, the evaluation was in 3D. Unless specified, we adopted these metric's notations throughout the paper. We also ran a statistical test using the Wilcoxon test \cite{whitley2002statistics} and we consider that the results are significantly different when the p-value is less than 0.05 (\ie~$p<0.05$). Moreover, we visually verify the plausibility of the results. In this context, plausibility refers to the complete, realistic, and anatomically possible segmentation. Therefore, a plausible or realistic output of a method contains no hole, no segment from unexpected regions of the image, and segmented areas are similar in shape to the targets’ structure.

\subsection{Prostate segmentation in radiotherapy}
\label{sec:TRUS}
Precise prostate gland segmentation from CT images is critical in the treatment of prostate cancer. However, it is not easy due to the inherently low contrast characteristics of CT for soft tissues and the appearance of high metal artifacts. For example, in CT exams of patients who have received low dose rate brachytherapy treatment for prostate cancer, in which tiny radioactive elements are permanently implanted into the gland, accurate prostate segmentation is difficult. The quality of the CT image is worsened by the high metal artifacts coming from the implanted radioactive elements. In our study, a clinical database of 78 CT prostate image cases was collected. All cases had received a prostate cancer treatment using a low dose rate brachytherapy technique. The in-plane resolution varies from $0.4$ $mm$ x $0.4$ $mm$ to $0.58$ $mm$ x $0.58$ $mm$ with a slice thickness between $1.5$ $mm$ and $2.5$ $mm$. We changed the datasets into the same voxel size of $0.5$ $mm$ x $0.5$ $mm$ x $1.25$ $mm$. A radiation oncologist with an experience of more than ten years ($>100$ implants/year) manually delineated the prostate. These delineation procedures are routinely done in the clinical treatment where it involves permanent brachytherapy with $^{125}I$ for localized prostate cancer \cite{davis2012american} \cite{girum2018inferring}. The dataset was randomly divided into 70\% for training, and the rest 30\% for testing and validation (20\% for testing and 10\% for validation). 
\begin{table*}[ht]
	\centering
	\caption{Results for a single target organ segmentation (mean $\pm$ standard deviation). All metrics are in 3D. The higher value of DSC is better, whereas the lower value is better for HD and RVD metrics (\textbf{bold} values are better).} 
	\label{tab:ALL}       
	\setlength{\tabcolsep}{14pt} 
	\renewcommand{\arraystretch}{1}
			\begin{tabular}{  l  l l  l l l  l l} 
				\hline
				{} & {}  &\multicolumn{6}{c}{Method} \\  
				\cline{3-8}
				Organ & Metric & FCN \cite{noh2015learning} & U-Net \cite{ronneberger2015u} & ResU-Net \cite{he2016deep} & Post-DAE \cite{larrazabal2020post}   & AGN \cite{schlemper2019attention} & \textbf{LFB-Net} \\
				\hline

				Prostate  &  DSC & 0.89$\pm$2.20  & 0.89$\pm$0.02 & 0.89$\pm$0.02 & 0.89$\pm$0.02 & 0.89$\pm$0.02 & \textbf{0.91$\pm$0.02}  \\

				(3D) & HD & 14.0$\pm$8.14 & 17.5$\pm$13.26 & 14.4$\pm$3.30 &11.3$\pm$3.30 &  16.6$\pm$7.12  & \textbf{6.1$\pm$1.40}  \\ 
				\hline
				Ear  & DSC& 0.93$\pm$0.05 & 0.92$\pm$0.06 &  0.93$\pm$0.05 &0.87$\pm$0.07 &0.93$\pm$0.05& \textbf{0.96$\pm$0.01}\\
				
				(3D)  & RVD & 0.07$\pm$0.08 & 0.08$\pm$0.09 &  0.06$\pm$0.09 &0.09$\pm$0.11 &0.04$\pm$0.05 & \textbf{0.02$\pm$0.02}\\		   
				\hline
			\end{tabular}
\end{table*}

From the experimental result shown in  Table \ref{tab:ALL}, our method can segment the prostate gland with an average volumetric Dice index of 0.91 and with a 3D HD of 6.1 mm. It appeared to outperform the other methods. For example, it produced a 2\% increased Dice index and decreased the 3D HD value by 11 mm over U-Net \cite{ronneberger2015u}. Our method without the feedback loop, \ie~only the forward system (FS), produces a Dice index of 0.90 and a 3D HD value of 10.3 mm. It improved both metrics over the other methods. Using the feedback loop reduces the 3D HD error value by 4.3 mm over not using it (only FS). Attention gated-based method \cite{schlemper2019attention} has only reduced the 3D HD metric over the U-net \cite{ronneberger2015u} by 0.9 mm, while the Post-DAE method \cite{larrazabal2020post} reduced it by  6.2 mm. However, the Post-DAE did not improve the average Dice index. The proposed LFB-Net method significantly outperformed $(p<0.05)$ the other methods for the Dice index value and 3D HD.

\subsection{Inner ear segmentation}
\label{inner_ear_segmentation}
Inner ear segmentation is essential for 3D visualization and modeling of the inner ear for surgeries. We chose the Hear-EU cochlear data descriptor public dataset consisting of $ \mu$CT-scans of 17 dry temporal bone specimens \cite{gerber2017multiscale}. The ground truth labels include the cochlear scala, semicircular canals and the vestibule. The images were acquired at 16.3 and 19.5 $ \mu$m voxel resolutions for 13 and 4 $ \mu$CTs, respectively. The original volume size of the $\mu$CTs ranged from 618 x 892 x 600 to 1500 x 1500 x 1500. The $\mu$CTs were standardized to a fixed size of 256 x 256 x 256 voxels for computational and memory requirements. The ground truth labels of 5 patients were not aligned with the $\mu$CT images. The data for these patients were manually corrected to align with the $ \mu$CT data perfectly. To compare all methods, a two-fold cross-validation technique was performed by randomly dividing the data at each fold into 76\% for the training and 24\% for the validation.  

As can be seen from Table \ref{tab:ALL}, LFB-Net achieved an increase in 4\% of the average Dice index and reduced relative volume difference by 6\% over the U-Net \cite{ronneberger2015u}. It also yielded accurate segmentation results without missing the unconnected small parts of the inner ear on CT images, which can then be used to model the inner ear structures for ear surgeries \cite{hussain2020augmented}. In contrast, the other methods yielded reduced segmentation results. In particular, the post-processing method, Post-DAE \cite{larrazabal2020post}, appeared to degrade the segmentation results of the U-Net \cite{ronneberger2015u}, in particular the Dice index by 5\%.
\subsection{Cardiac cine-MRI segmentation}
The third application of our method, which aims to demonstrate the performance of our model for multi-class segmentation applications, is on cardiac cine-MRI segmentation. Although short-axis cardiac cine-MRI is essential to study the cardiac function of the left and right ventricles, accurate segmentation of both ventricles remains challenging. The segmentation of the endocardial border in diastole and systole is required to evaluate the cardiac function (\ie~cavity volume and ejection fraction) for the left and right ventricles, and the epicardial border of the left ventricle is mandatory to evaluate the myocardial mass and thickness. The main challenge is the variability in the shape of the left and right ventricular cavities. In this regard, we chose the Automated Cardiac Diagnosis challenge (ACDC) \cite{bernard2018deep} to evaluate the proposed method. It contains 100 patients for the training and 50 for the testing. These datasets are available for download at \footnote{https://acdc.creatis.insa-lyon.fr/}. 
To analyze how the baseline methods would work in the testing sets, we have access to the ground truth upon request to the organizers.
We randomly divided the 100 cases into 75 patients (75\%) for training and 25 patients (25\%) for validation (all the cases with diastole and systole phases). Thus, to further inspect the performance of the method, we evaluated the end-diastolic and end-systolic phases separately on the 50 testing cases.

In all 3D experimental measurements on the validation sets shown in Table \ref{tab:Cardiac-MRI_camus_validation segmentation}, our method appeared to improve segmentation accuracy. It yielded an average Dice and HD values, respectively of $0.96\pm0.02$ and $6.7\pm3.95$ mm for the left ventricular cavity (LV), of $0.91\pm0.05$ and $13.2\pm7.90$ mm for the right ventricular cavity (RV), and of $0.90\pm0.03$ and $9.4\pm5.21$ mm for the myocardium (MYO) on the validation datasets. The 3D HD value of the myocardium refers to the largest distance error from either the endocardium or the epicardium. More importantly, we observed that the improvement is significant in cases where the segmentation task is difficult, such as in the right ventricular cavity and the myocardium. 

As shown in Table \ref{tab:Cardiac-MRI segmentation_testing_resuls}, on the ACDC testing set, our method yielded an average Dice index and 3D HD values, respectively of $0.94\pm0.06$ and $7.5\pm5.54$ mm in the LV, of $0.92\pm0.06$ and $11.9\pm6.49$ mm in the RV, and $0.90\pm0.03$ and $9.5\pm5.58$ mm in the MYO. We observed that LFB-Net significantly outperforms the other methods ($p < 0.05$). To further analyze the method’s generalizability property, results from the unseen testing set and validation set are shown in Table \ref{tab:tola_acdc_segmentation_results}. Although we have different sample sizes for the validation set, 25 cases (each with end-diastolic and end-systolic phases), and the testing set, 50 cases  (each with end-diastolic and end-systolic phases),  these average values in Table \ref{tab:tola_acdc_segmentation_results} can be considered to infer the methods’ performance for each set.
Indeed, the proposed method showed almost no differences between the unseen testing and the validation sets by achieving a total Dice index of 0.92 and a 3D HD value of around 10 $mm$. In contrast, the other methods showed a large difference between the validation and testing sets. For example, LFB-Net has significantly outperformed the other methods in the heart segmentation on the testing data (shown in Table \ref{tab:tola_acdc_segmentation_results}). However, this is not always the case on the validation data.

\begin{table*}[ht]
	\centering
	\small
	\caption{Multi-structure cardiac image segmentation results of the \textbf{validation set} for end-diastolic (ED) and end-systolic (ES) phases. LV: Left ventricular cavity; MYO: Myocardium, RV/LA: Right ventricular cavity (RV) for the Cine-MRI and left atrium cavity (LA) for the echocardiography. The bold values refer to the best performance for each metric.}
	\label{tab:Cardiac-MRI_camus_validation segmentation}       
		\setlength{\tabcolsep}{7pt} 
		\begin{tabular}{l l l l l l l l l l l l l l} 
			\hline
			 & {}	&\multicolumn{4}{c}{LV}& \multicolumn{4}{c}{RV/LA} & \multicolumn{4}{c}{MYO} \\
			\cline{3-5}
			\cline{7-9}
			\cline{11-13}
			& {}&\multicolumn{2}{c}{\underline{Dice}}&  \multicolumn{2}{c}{\underline{HD}} & \multicolumn{2}{c}{\underline{Dice}} &  \multicolumn{2}{c}{\underline{HD}} & \multicolumn{2}{c}{\underline{Dice}} &  \multicolumn{2}{c}{\underline{HD}}\\ 
			Data & Method &\multicolumn{1}{c}{ES}&  \multicolumn{1}{c}{ED} & \multicolumn{1}{c}{ES}&  \multicolumn{1}{c}{ED} &\multicolumn{1}{c}{ES}&  \multicolumn{1}{c}{ED} &  \multicolumn{1}{c}{ES}&  \multicolumn{1}{c}{ED}  &\multicolumn{1}{c}{ES}&  \multicolumn{1}{c}{ED} & \multicolumn{1}{c}{ES}&  \multicolumn{1}{c}{ED}\\
			\hline
	Echocardiography (2D)   & AGN \cite{schlemper2019attention} & 0.92 &0.95 &9.7 &6.0 & 0.92 & 0.89 &7.1 & 8.5 &0.69 &0.69 &24.2  &25.0
	\\
	
	& FCN \cite{noh2015learning} &0.92 &0.95 &5.5 &5.7 &0.92 &0.90 &4.8  &5.8 &0.86 &0.85 & 7.0 &8.1 \\
		& U-Net \cite{ronneberger2015u}&0.93 &0.95 &5.4  &\textbf{5.4} &0.92 &0.89 &5.1  &6.5 &\textbf{0.87} &0.86 &7.4  &8.3 \\
	
	  & ResU-Net \cite{he2016deep} &0.92 &0.94 &5.7 &6.0 &0.91 &0.88 &5.4  &6.3 &0.86 &0.86 &7.6  &8.4 \\
	 
	(54 cases) & \textbf{LFB-Net} & \textbf{0.93} &\textbf{0.95} &\textbf{5.3} &\textbf{5.4} &\textbf{0.93} &\textbf{0.91} &\textbf{4.6}  &\textbf{5.4} &\textbf{0.87} &\textbf{0.87} &\textbf{6.7}  &\textbf{7.1} \\			
			\hline
	Cine-MRI (3D) & AGN \cite{schlemper2019attention} & 0.94 &0.96 &7.8 &6.9 & 0.84 & 0.93 &15.1 & 14.3 &0.90 &0.89 &11.1  &10.1 \\
	
	& FCN \cite{noh2015learning} &0.93  &0.96  &7.8 &6.7  &0.84 &0.92 &15.58 &14.7 &0.89&0.87 &11.1 &10.0\\
		& U-Net \cite{ronneberger2015u} & 0.93  &0.96  &8.3  &8.3 & 0.84 &0.93 &16.2 & 14.4 & 0.90  &0.88  &11.9  &9.6 \\
	& ResU-Net \cite{he2016deep} &0.92 &0.95  &9.3  &8.5 & 0.85 &0.93 &15.5 &15.9 &0.89  & 0.88 &11.7  & 10.4\\


	 (25 cases) & \textbf{LFB-Net} & \textbf{0.94}  & \textbf{0.97} &\textbf{6.7}  &\textbf{6.5} & \textbf{0.88} & \textbf{0.95} &\textbf{13.7} &\textbf{12.9} & \textbf{0.91} &\textbf{0.89} &\textbf{10.5}&\textbf{8.4} \\	
	\hline
		\end{tabular}
\end{table*}

\subsection{Echocardiographic image segmentation}
Accurate cardiac structure segmentation from echocardiography images is profoundly vital in cardiac diagnosis. For this, we chose Multi-structure Ultrasound Segmentation (CAMUS)
dataset to evaluate our method \cite{leclerc2019deep}. It contains two and four-chamber acquisition from 500 patients, with end-diastolic and end-systolic phases. Thus, a given patient has four images (two in end-diastole and two in end-systole). The segmentation reference and raw data of 450 patients are available for download at \footnote{https://camus.creatis.insa-lyon.fr/challenge/}. We randomly divided this data into 396 patients for training and 54 patients for validation.

As shown in Table \ref{tab:Cardiac-MRI_camus_validation segmentation}, on the 54 validation exams, our method has improved both the Dice and the HD values. For the two view acquisition (two-chambers and four-chambers) based segmentation results, our method yielded an average Dice index of $0.94\pm0.03$, $0.92\pm0.04$, and $0.86\pm0.06$ for the 4-chamber,  and $0.94\pm0.03$, $0.92\pm0.05$, and $0.88\pm0.04$ for the 2-chamber respectively in the left ventricular cavity (LV), left atrium (LA), and myocardium (MYO). The average HD values were $5.0\pm2.83$ mm, $5.2\pm3.48$ mm, and $6.7\pm3.04$ mm for the 4-chambers, and $5.6\pm3.22$ mm, $4.8\pm2.79$ mm, and $7.1\pm3.86$ mm for the 2-chambers respectively in the LV, LA, and MYO. Although it produces similar results for the different view acquisitions in LV and LA, it improved the myocardium segmentation on the 2-chambers over the 4-chambers by 2\% in the Dice index.

\begin{table*}[ht]
\centering
	\caption{Results for cardiac cine-MRI segmentation on the ACDC \textbf{testing set (50 cases)} at end-diastolic (ED) and end-systolic (ES) phases. LV: Left ventricular cavity; MYO: Myocardium, RV: Right ventricular cavity (RV). * $(p<0.05)$ indicates that the difference between LFB-Net and the other methods is significant. Values are expressed as a mean $\pm$ standard deviation in 3D.}
	\label{tab:Cardiac-MRI segmentation_testing_resuls}        
	\small
		\setlength{\tabcolsep}{5pt} %
		\begin{tabular}{l l l l l l l l l l l l l} 
 \toprule 
		 {}	&\multicolumn{4}{c}{LV}& \multicolumn{4}{c}{RV} & \multicolumn{4}{c}{MYO} \\
			\cline{2-4}
			\cline{6-8}
			\cline{10-12}
			{}&\multicolumn{2}{c}{\underline{Dice}}&  \multicolumn{2}{c}{\underline{HD}} & \multicolumn{2}{c}{\underline{Dice}} &  \multicolumn{2}{c}{\underline{HD}} & \multicolumn{2}{c}{\underline{Dice}} &  \multicolumn{2}{c}{\underline{HD}}\\ 
			Method &\multicolumn{1}{c}{ES}&  \multicolumn{1}{c}{ED} & \multicolumn{1}{c}{ES}&  \multicolumn{1}{c}{ED} &\multicolumn{1}{c}{ES}&  \multicolumn{1}{c}{ED} &  \multicolumn{1}{c}{ES}&  \multicolumn{1}{c}{ED}  &\multicolumn{1}{c}{ES}&  \multicolumn{1}{c}{ED} & \multicolumn{1}{c}{ES}&  \multicolumn{1}{c}{ED}\\
			\hline
	AGN \cite{schlemper2019attention} & 0.89&0.96 & 11.5&7.9 & 0.83 & 0.91 & 16.5& 13.4&0.89&0.88&12.5&10.5\\
	
	& $\pm$0.10*  & $\pm$0.02* & $\pm$9.73* & $\pm$5.37* &$\pm$0.14* & $\pm$0.09* &$\pm$10.39* &$\pm$8.33* &$\pm$0.05* &$\pm$0.04* &$\pm$5.69* &$\pm$6.12* \\
			
	FCN \cite{noh2015learning} &0.89 &0.96   &10.8  & 7.6
&0.85  &0.90  &15.8  &14.0 &0.87  &0.86  &12.6  &11.3\\
& $\pm$0.09*   &$\pm$0.02*&$\pm$6.2*  &$\pm$4.49* &$\pm$0.12*  &$\pm$0.07*  &$\pm$7.70*  &$\pm$6.55* &$\pm$0.05*  &$\pm$0.04*  &$\pm$4.60*  &$\pm$6.09*\\

 U-Net \cite{ronneberger2015u} &0.89  &0.96 &11.3  &8.2
&0.83  &0.90  &16.8 &14.3
&0.89 &0.87  &12.5  &10.9\\

&$\pm$0.09*   &$\pm$0.02*   &$\pm$7.90*  &$\pm$5.0* &$\pm$0.17* &$\pm$0.12* &$\pm$10.2*  &$\pm$8.29* &$\pm$0.04* &$\pm$0.04*  &$\pm$5.1*  &$\pm$6.27* \\

ResU-Net \cite{he2016deep} &0.90  &0.96  &10.5  &8.6
&0.85 &0.90 &15.2  &13.5
&0.88 &0.87 &12.6  &10.8\\
&$\pm$0.10*&$\pm$0.02* &$\pm$7.34* &$\pm$5.0* &$\pm$0.11*  &$\pm$0.12* &$\pm$6.91* &$\pm$8.22* &$\pm$0.09* &$\pm$0.04* &$\pm$7.98* &$\pm$6.39*\\

\textbf{LFB-Net} &\textbf{0.92}   &\textbf{0.97} &\textbf{8.5} &\textbf{6.5}
&\textbf{0.89}  &\textbf{0.94}  &\textbf{13.0}  &\textbf{10.9}
&\textbf{0.91}   &\textbf{0.89}  &\textbf{9.9}  &\textbf{9.1} \\
&\textbf{$\pm$0.07}   &\textbf{$\pm$0.02} &\textbf{$\pm$6.76} &\textbf{$\pm$3.84} &\textbf{$\pm$0.08}  &\textbf{$\pm$0.04}  &\textbf{$\pm$6.46}  &\textbf{$\pm$6.49} &\textbf{$\pm$0.03}   &\textbf{$\pm$0.03}  &\textbf{$\pm$5.33}  &\textbf{$\pm$5.92} \\

\bottomrule
\end{tabular}
\end{table*}

\begin{table}[ht]
\caption{Total heart segmentation performance on ACDC validation (Valid.) and testing (Test.) sets in 3D by computing the average values of the MYO, LV, and RV segmentation. * $(p<0.05)$ indicates that the difference between LFB-Net and the other methods is significant.}
	\label{tab:tola_acdc_segmentation_results}       
	\setlength{\tabcolsep}{5pt} 
	\renewcommand{\arraystretch}{1}
	\begin{tabular}{ l  l l l l } 
		\hline
		{}  &  \multicolumn{4}{c}{Metric}  
		\\    
		\cline{2-5}
		Method &\multicolumn{2}{c}{\underline{Dice}} & \multicolumn{2}{c}{\underline{HD (mm)}}\\
		
	&  \hspace{0.5cm}Valid.&  \hspace{0.5cm}Test. & \hspace{0.5cm}Valid. &\hspace{0.5cm} Test.\\
		\hline
	AGN \cite{schlemper2019attention} & 0.910$\pm$0.07*& 0.894$\pm$0.09* &10.9$\pm$7.13 & 12.0$\pm$8.23* \\ 
	
	FCN \cite{noh2015learning} &  0.903$\pm$0.07* & 0.890$\pm$0.08* & 11.0$\pm$7.22 & 12.0$\pm$6.53*
		\\
		
	U-Net \cite{ronneberger2015u} & 0.906$\pm$0.06* & 0.888$\pm$0.10* & 11.5$\pm$7.50 & 12.3$\pm$7.79*\\

		ResUnet \cite{he2016deep} & 0.903$\pm$0.07& 0.893$\pm$0.09* & 11.9$\pm$8.39*& 11.9$\pm$7.32*\\		
		
		\textbf{LFB-Net} &  \textbf{0.921$\pm$0.05} & \textbf{0.920$\pm$0.06} & \textbf{9.9$\pm$6.47} &\textbf{9.7$\pm$6.18}\\
		\hline
	\end{tabular}
\end{table}

\begin{table*}[ht]
\centering
	\caption{Long-axis echocardiographic image segmentation results on the \textbf{testing} CAMUS data at end-diastolic (ED) and end-systolic (ES) phases. Comparisons are shown for our LFB-Net method and CAMUS challengers. Results were obtained from the CAMUS challenge portal. The provided inter- and intra-observer values were from only 40 cases (good and medium
image qualities) by excluding ten low-quality image cases. No inter- and intra-observer studies were provided for the left atrium \cite{leclerc2019deep}.}
	\label{tab:echocardiographic image segmentation}      
	\setlength{\tabcolsep}{5.7pt} 
		\begin{tabular}{l l l l l l l l l l l l l l l} 
\toprule
			 & & {}	&\multicolumn{4}{l}{LV: Endocardium}& \multicolumn{4}{l}{LV: Epicardium} & \multicolumn{4}{l}{Left Atrium} \\
			\cmidrule(lr){4-7}
			\cmidrule(lr){8-11}
			\cmidrule(lr){12-15}
			& & {}&\multicolumn{2}{c}{\underline{Dice}}&  \multicolumn{2}{c}{\underline{HD}} & \multicolumn{2}{c}{\underline{Dice}} &  \multicolumn{2}{c}{\underline{HD}} & \multicolumn{2}{c}{\underline{Dice}} &  \multicolumn{2}{c}{\underline{HD}}\\ 
			Data &  & Method &\multicolumn{1}{c}{ES}&  \multicolumn{1}{c}{ED} & \multicolumn{1}{c}{ES}&  \multicolumn{1}{c}{ED} &\multicolumn{1}{c}{ES}&  \multicolumn{1}{c}{ED} &  \multicolumn{1}{c}{ES}&  \multicolumn{1}{c}{ED}  &\multicolumn{1}{c}{ES}&  \multicolumn{1}{c}{ED} & \multicolumn{1}{c}{ES}&  \multicolumn{1}{c}{ED}\\
			\hline
			
			

	Echo-& & inter-observer & 0.873 & 0.919&6.6&6.0&0.890& 0.913& 8.6& 8.0& - & - & -&- \\
	cardiography& &  & $\pm$0.060 & $\pm$0.033 &$\pm$2.4 &$\pm$2.0 & $\pm$0.047 & $\pm$0.037 & $\pm$3.3 & $\pm$ 2.9& - & - & -&-\\

	(2D)& & Intra-observer & 0.930& 0.945& 4.5& 4.6& 0.951& 0.957& 5.0& 5.0& - & - & - & -\\
		& &  & $\pm$0.031 & $\pm$0.019 &$\pm$1.8 &$\pm$1.8 & $\pm$0.021 & $\pm$0.019 & $\pm$2.1 & $\pm$2.3 & - & -&-&-\\
		
	\cmidrule(lr){3-15}
	& & Oktay O. \etal~ \cite{oktay2017anatomically}&0.913 &0.936 &5.6 & 5.6 & 0.945  &0.953 &5.9 &5.9 & 0.911  &0.881 &5.8 & 6.0\\
	& & Leclerc S. \etal~ \cite{leclerc2019deep} &0.912  &0.936 & 5.5 &5.3 &0.946 & 0.956 &5.7 & \textbf{5.2} & 0.918  &0.889 &5.3 & 5.7\\
			
	Testing  & & U-net-2 \cite{leclerc2019deep}  &0.899  &0.922 &5.3 & 5.7 &0.923  &0.932 &6.4 & 6.4 &0.888 &0.848 &6.2 &6.9\\
	   (50 cases) & & \textbf{LFB-Net} &\textbf{0.926}  &\textbf{0.946} &\textbf{4.8} &\textbf{4.8} &  \textbf{0.952}  &\textbf{0.959} &\textbf{5.2} &\textbf{5.2}&\textbf{0.924}  & \textbf{0.902} &\textbf{5.0} &\textbf{5.2}\\	\bottomrule
		\end{tabular}
\end{table*}

As shown in Table \ref{tab:echocardiographic image segmentation}, on the 50 CAMUS testing exams, LFB-Net achieved an average Dice index of $0.96\pm0.02$ for the LV epicardium, $0.94\pm0.03$ for the LV endocardium, and $0.91\pm0.07$ for the left atrium. It outperformed the other CAMUS challengers \cite{oktay2017anatomically} \cite{leclerc2019deep}. LFB-Net improves the segmentation in all multi-label structures with less accuracy variability among the test data. Moreover, it  notably improved the segmentation by large value on the left atrium in the end-diastolic phase. However, as the results were obtained by submitting the predicted images to the challenge website, we could not perform a statistical comparison \cite{leclerc2019deep}. The proposed method also achieved comparable results with the intra-observer values. Mainly, it yielded better results in the Dice index except for the endocardium in the systolic  phase. Thus, segmentation with the context feedback loop yields consistent results.
\subsection{Qualitative segmentation results}
\label{sec:qualitative_results}
\begin{figure}[h]
	\centering
	\includegraphics[width=1\linewidth]{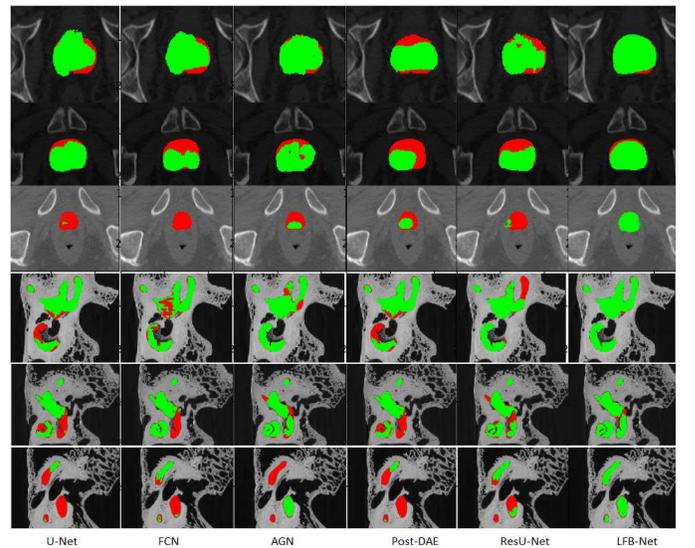}
	\caption{Examples of single label segmentation results. [Rows 1-3] Prostate segmentation examples, and [Rows 4-6] inner ear segmentation examples. The predicted mask (in green) is overlapped with the ground truth (in red).}	
	\label{fig_single_label_qualitative results_sin}
\end{figure}

\begin{figure}[ht]
	\centering
	\includegraphics[width=1\linewidth]{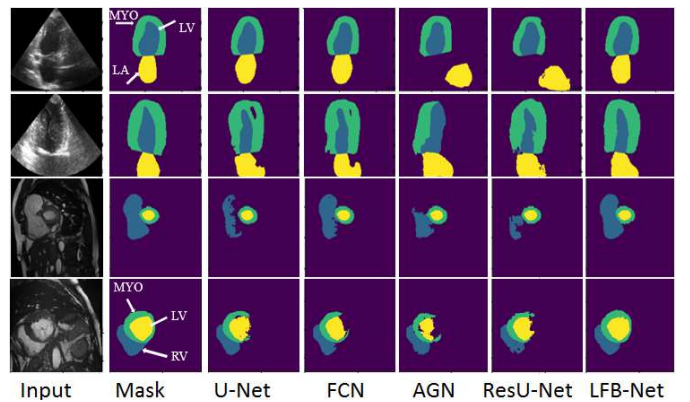}
	\caption{Examples of multi-label cardiac image segmentation results. [Row 1-2] long-axis echocardiography, and [Row 3-4] short axis cine-MRI segmentation. RV: Right ventricular cavity; LV: Left ventricular cavity; MYO: Myocardium; and LA: Left atrium.}	
	\label{fig_multi_label_qualitative results}
\end{figure}

As one can observe from the qualitative segmentation results in Fig. \ref{fig_single_label_qualitative results_sin} for the single label and Fig. \ref{fig_multi_label_qualitative results} for the multi-label segmentation, our model produces more plausible results than the other methods. From a careful visual checking of the results, if it is with holes for a given single structure segmentation and between structures for multi-structure segmentation and comparing the shape, our method produces more plausible results. Whereas the other methods produce holes in a given structure or between structures and sometimes produce atypical results which are type of errors that could not be made by manual segmentation. Moreover, as seen in the ear segmentation [Row 4-6], the other methods appeared to fail in scenarios when they segment multiple unconnected small structures. In contrast, our method produces a more realistic segmentation of all structures. We observed similar scenarios throughout the testing data. Indeed, anatomical plausibly is a prerequisite for the experts to use the segmented structures for clinical assessments. With the proposed method, the reliability of the segmentation renders trustworthy the clinical information extracted from these segmented structures.

\subsection{Network ablation study}
\label{sec:abalation_study}
\subsubsection*{\textbf{Ablation study for system design}}
To evaluate the contribution of each building block in our method, 
we created the following configurations. 
\begin{enumerate}
	\item The forward system (FS): a U-Net architecture without the feedback looping system in both the training and the testing phase (using only step 1 in Fig. \ref{fig_training}). 
	\item The FS*: a forward system (FS) without the squeeze-and-excitation network, to study its effect in our method.  	
	\item The proposed method (LFB-net): Forward system regularized by an FCN-based feedback system during the training and the testing phases.
\end{enumerate}

\begin{figure}[ht]
	\centering
	\includegraphics[width=1.02\linewidth]{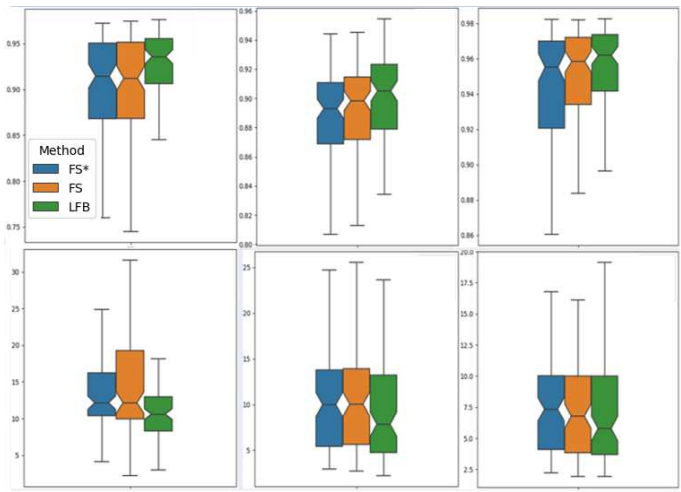}
	\caption{Box plot of the ablation study for system design on the ACDC validation set. [row 1] Mean Dice coefficient (higher is better),  [Row 2] mean 3D HD (mm) (lower is better). [Column 1] RV, [column 2] MYO, and [column 3] LV. FS: Forward system; FS*: Forward system without the squeeze-and-excitation network; and LFB: the proposed final method.}
	\label{fig_ablation}
\end{figure}

Figure \ref{fig_ablation} shows the ablation study on the 25 ACDC validation cases. It can be seen that training with the feedback loop consistently increased the results. 
Moreover, considering the feedback system's encoder during the testing has improved the accuracy of using it only during the training. As shown in Fig. \ref{fig_ablation}, our method produces less inter case difference, yielding smaller standard deviations in both Dice and 3D HD metrics. The Se-block \cite{hu2018squeeze} has also appeared to increase the forward system's accuracy, yielding better results than without it. It is always true in the total average values. We found that segmentation with the feedback loop significantly outperforms the other two network configurations (FS and FS*) in both Dice and 3D HD metrics of the RV and MYO ($p <0.05$). However, although the feedback loop has improved the average performance for LV segmentation, it was not significant. Note that most networks, including the SOTA methods, performed well for LV segmentation but not for the MYO and RV segmentation.  Thus, these results guarantee us to say that using the feedback loop increases the segmentation accuracy, but significantly for the complicated structures. 

\begin{figure}[ht]
	\centering
	\includegraphics[width=1\linewidth]{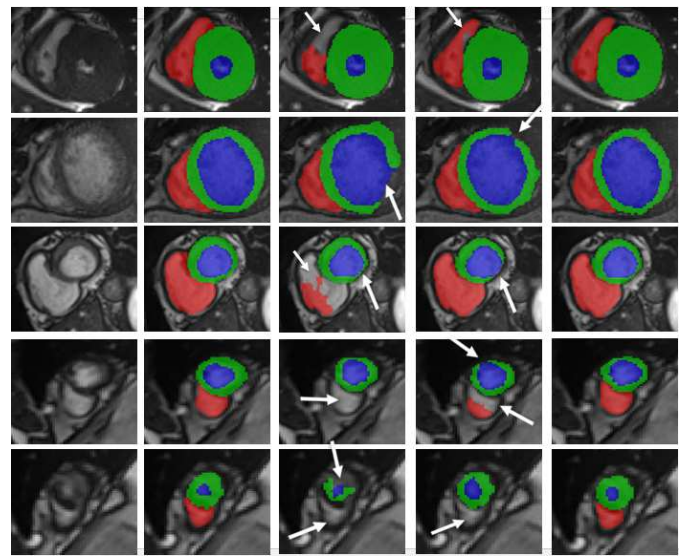}
	\caption{Examples of the ablation study for system design on the ACDC data. From left to right: original image, ground truth, FS, FS*, and our method (LFB-Net). [rows 1-5] Cases where the feedback loop based segmentation (\ie~LFB-Net) corrects wrong segmentation from the forward system. The red, green, and blue colors show the right ventricular cavity, the myocardium, and the left ventricular cavity, respectively. The white arrows show the areas where FS and FS* fails to segment the targets accurately.}
	\label{fig_qualitative_ablataion}
\end{figure}

As shown in Fig. \ref{fig_qualitative_ablataion}, training with the feedback loop mitigates the conditions in which the forward system potentially fails. Firstly, in these examples, the channel-wise feature calibration (i.e., squeeze-and-excitation network) improves the segmentation over using only convolutional networks. It is consistent with the quantitative results in Fig. \ref{fig_ablation}. Training with the feedback loop improves the forward system's accuracy, primarily when it produces low quality image labels. On these examples, we observed that the improvement was mainly at the basal and apical regions of the heart, in particular for the right ventricular cavity segmentation at the end-systolic phase, and the endocardial and epicardial border segmentation at the end-diastolic phase. These qualitative results are consistent with the quantitative results presented in Tables \ref{tab:Cardiac-MRI segmentation_testing_resuls} and \ref{tab:echocardiographic image segmentation} for the ACDC and CAMUS testing datasets, respectively. Most other methods produce less good quantitative results for structures that are difficult to segment, such as the right ventricular cavity, the endocardial and epicardial borders for the ACDC data, and the left atrium for the CAMUS data. In contrast, LFB-Net produces better results for these regions. 

The proposed feedback system is integrated with the forward system while training. The optimal trained neural network weights of the forward and feedback systems' are thus saved simultaneously as final models. This design enables the forward system to always benefit from the feedback loop. In contrast, this might not always be the case while using post-processing methods \cite{larrazabal2020post} \cite{painchaud2020cardiac}. For example, the denoising autoencoder-based method \cite{larrazabal2020post} degraded the Dice index of Ronneberger \etal \cite{ronneberger2015u} in the inner ear segmentation by 5\%.

To further study the hypothesis that feedback loop would increase accuracy, specifically for difficult or noisy images, we calculated the percentage of the performance of the methods on CAMUS datasets by specifying threshold values of whose Dice is less than 0.88 or HD error is greater than 6.5 mm. Experimental results are shown in Table \ref{tab:abalation_percentage}, indicating that most results of our method are above 0.88 Dice and below 6.5 HD error, outperforming the other methods. It is particularly true for the MYO and LA. However, it yielded a 2.8\% lower HD value than the U-Net in LV \cite{ronneberger2015u}, which is not a difficult structure for segmentation. These results further demonstrate that the feedback loop has a significant benefit to segment difficult structures such as MYO and LA.

\begin{table}[ht]
	\centering
	\small
	\caption{Percentage of cases whose Dice is below 88\% or HD is above 6.5 mm for 54 CAMUS validation sets.}
	\label{tab:abalation_percentage}       
	\setlength{\tabcolsep}{4pt} 
	\renewcommand{\arraystretch}{1}
	\begin{tabular}{ l   l l l | l l l } 
		
 \toprule 		 
		 & \multicolumn{3}{c}{Dice}
		&\multicolumn{3}{c}{HD (mm) } \\
		& \multicolumn{3}{c}{\% (minimum value)}
		&\multicolumn{3}{c}{\% (maximum value)} \\
		\hline 
		Method& MYO & LA  & LV & MYO & LA & LV\\
		\hline
	FCN   & 53.9\%&16.4\%&5.6\%  & 51.4\% & 22\% & 29.4\% \\
	&(0.54) & (0.36) & (0.74) &(34.4)&\textbf{(19.6)}&(19.9)  \\
	\hline 

U-Net & 49.5\% & 17.8\% & 6.1\% & 50.9 \% & 24.3\%& \textbf{24.3\% }  \\
  &(0.56)& (0.64) & (0.72)) & (39.2)&(46.6)&(20.6 \\
\hline
ResU-Net & 56.1\% & 17.8\% & 8.9\% & 54.2\% & 25.7\% & 34.1\%  \\
& (0.60)&(0.53)&(0.79) & (54.2)& (25.7) & (34.1)\\
\hline 
AGN & 95.8\% & 20.1\% & 6.5\% & 96.7\% & 34.1\% & 24.3\% \\
&(0.18) & (0.5) &(0.73)& (93.5) & (126.8) & (89.0)\\
\hline
\textbf{LFB-Net}& \textbf{44.4\%}& \textbf{13.1}\% &\textbf{4.7\%} & \textbf{46.3\%} & \textbf{18.7\%} & 27.1\%  \\

& \textbf{(0.65)}&\textbf{(0.74)}&\textbf{(0.80)} &\textbf{(30.6)}&\textbf{(19.6)} & \textbf{(16.0)}\\
\bottomrule
	\end{tabular}
\end{table}
To examine where our method best performs, we computed the maximum HD error and the minimum Dice coefficient. Lower Dice and higher HD reveal the model's worst scores. Results are shown in Table \ref{tab:abalation_percentage}, illustrating that our method considerably decreased the maximum errors in every metric.

Another essential advantage of our approach is that it produces segmentation with almost no difference in the testing data populations. It can be observed from Table \ref{tab:ALL}, for prostate and inner ear segmentation, that the standard deviation is small in all measurements. It was similar for the cardiac segmentation from both cine-MRI and echocardiographic images.

\subsubsection*{\textbf{Ablation study for system integration strategy}}
To investigate the best combination strategy of the two systems through their latent spaces (\ie~$h_{s}$ and $h_{f}$), we performed three different schemes such as concatenation, addition, and multiplication on the testing prostate datasets. 
 
\begin{figure}[ht]
		\centering
		\includegraphics[width=0.9\linewidth]{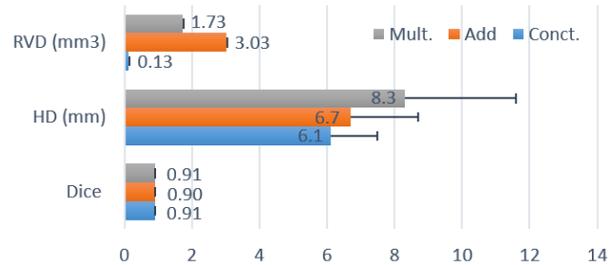}
	\caption{Merging strategy comparison. Combination strategies (Concat: concatenation, Add: Addition, Multi: Multiplication, \ie~merge $h_{s}$ and $h_{f}$). 
				Best performance is with higher Dice, and with lower 3D HD and RVD values.}
		\label{fig_ablation_merging_strategy}
\end{figure}

As shown in Fig. \ref{fig_ablation_merging_strategy}, the concatenation strategy appears to outperform the other two strategies. Thus, we selected the concatenation layer
to merge the latent spaces. Indeed, statistically, the three strategies showed no significant difference for the Dice index, but in the 3D HD metric, the concatenation strategy significantly outperformed the others.   


\subsubsection*{\textbf{Ablation study for system training scheme}}
Our method is based on an alternative training strategy of a modified U-Net (forward system), and FCN architecture (feedback system). The FCN network is aiming to regularize the forward system by showing back its predicted probabilistic output, thereby improving the learning ability over time. For this, we conducted experiments on ACDC data to compare the alternative training schemes: with no feedback loop and with the feedback loop. 

\begin{figure}[ht]
\centering
\includegraphics[width=1\linewidth]{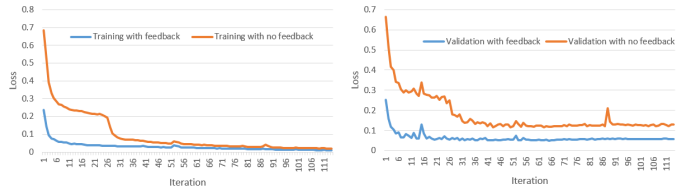}
\caption{Ablation study for the training scheme on ACDC data. The red line represents the forward system's training loss with no update of the decoder from the feedback loop. The blue represents the forward system's loss with the feedback loop $h_{f}$. Segmentation with the feedback loop accelerates the convergence speed with lower validation loss.}	
\label{fig_training_scheme}
\end{figure}

Figure \ref{fig_training_scheme} shows how the training loss changes in the two scenarios. Training and validation losses decrease faster while training with the feedback loop than training without the feedback loop. Besides, although the training loss without the feedback tends to decrease over time, validation loss is not. This further shows that the model without the feedback tends to quickly over-fit to the training data over iterations than the model with the feedback.  

Segmentation with the feedback loop took an extra 1 hour training time to converge than without the feedback loop on the ACDC data. Although we designed the segmentation problem as a two systems task, the total number of parameters needed to optimize is smaller than the one model-based methods. Our method has 8.5 and 7.9 million parameters during the training and the testing phases, respectively, which are computationally more efficient than the 32 million trainable parameters of U-Net architecture \cite{ronneberger2015u}. Thus, our method can deliver results quickly, which can be beneficial for real-time applications. For example, it produces a $256\times256\times4$ segmentation result (4 indicates the predicted probabilistic output for LV, RV, MYO, and background) of the cine-MRI within $0.025$s on a personal computer of i7 with 32 GB RAM. 
\section{Conclusions}
\label{sec:conclusion}
In this paper, we introduced a deep learning method for accurate and robust medical image segmentation by formulating the segmentation problem as a two systems task. It employs a forward system (modified U-Net) for hierarchal feature extraction-driven image segmentation along with a contextual feedback system. The FCN-based contextual feedback system is used to regulate the forward system's segmentation process. It allowed the forward system to attend and improve its previous decisions, particularly on the uncertain image regions over time. This strategy of modeling image segmentation as a two systems task enabled us to develop an efficient architecture that can be trained from a small dataset and quickly deliver segmentation results. 

We demonstrated our method's performance through extensive ablation and experimental results in the prostate, short-axis cardiac-MRI, inner ear, and long-axis echocardiographic image segmentation applications. The experimental results reveal two important points. Firstly, a spatial feedback loop-based image segmentation is an effective feed-forward learning approach that produces both plausible and accurate segmentation results. The plausibility was achieved without incorporating shape prior or applying post-processing method. Secondly, our method produces results with reduced segmentation variability among the testing data that shows robustness to segment low contrast images as well as structures. It also provides results with reduced maximum errors in all metrics. Moreover, the proposed method yielded significantly better results than the state of the art methods for single and multi-structure segmentation, specifically for the complex structures. Thus, our work opens important perspectives towards efficient and accurate medical image analysis tasks by interconnecting two networks through the introduced feedback loop method. 

Moreover, the proposed LFB-Net framework can be extended to other medical image analysis tasks. In this regard, future research will focus on exploring conditions on how to exploit the contextual feedback loop’s latent space and efficiently leverage the merged contextual information. In cases of available 3D datasets, the 3D version of the proposed method could be applied to capture the 3D topology of the target.
\bibliographystyle{IEEEtran}
\bibliography{bibtex/bib/IEEEabrv.bib}

\begin{thebibliography}{10}
\providecommand{\url}[1]{#1}
\csname url@samestyle\endcsname
\providecommand{\newblock}{\relax}
\providecommand{\bibinfo}[2]{#2}
\providecommand{\BIBentrySTDinterwordspacing}{\spaceskip=0pt\relax}
\providecommand{\BIBentryALTinterwordstretchfactor}{4}
\providecommand{\BIBentryALTinterwordspacing}{\spaceskip=\fontdimen2\font plus
\BIBentryALTinterwordstretchfactor\fontdimen3\font minus
  \fontdimen4\font\relax}
\providecommand{\BIBforeignlanguage}[2]{{%
\expandafter\ifx\csname l@#1\endcsname\relax
\typeout{** WARNING: IEEEtran.bst: No hyphenation pattern has been}%
\typeout{** loaded for the language `#1'. Using the pattern for}%
\typeout{** the default language instead.}%
\else
\language=\csname l@#1\endcsname
\fi
#2}}
\providecommand{\BIBdecl}{\relax}
\BIBdecl

\bibitem{davis2012american}
B.~J. Davis, E.~M. Horwitz, W.~R. Lee, J.~M. Crook, R.~G. Stock, G.~S. Merrick,
  W.~M. Butler, P.~D. Grimm, N.~N. Stone, L.~Potters \emph{et~al.}, ``American
  brachytherapy society consensus guidelines for transrectal ultrasound-guided
  permanent prostate brachytherapy,'' \emph{Brachytherapy}, vol.~11, no.~1, pp.
  6--19, 2012.

\bibitem{girum2018inferring}
K.~B. Girum, A.~Lalande, M.~Quivrin, I.~Bessi{\`e}res, N.~Pierrat, E.~Martin,
  L.~Cormier, A.~Petitfils, J.~M. Cosset, and G.~Cr{\'e}hange, ``Inferring
  postimplant dose distribution of salvage permanent prostate implant (ppi)
  after primary ppi on ct images,'' \emph{Brachytherapy}, vol.~17, no.~6, pp.
  866--873, 2018.

\bibitem{petitjean2011review}
C.~Petitjean and J.-N. Dacher, ``A review of segmentation methods in short axis
  cardiac mr images,'' \emph{Medical image analysis}, vol.~15, no.~2, pp.
  169--184, 2011.

\bibitem{bernard2018deep}
O.~Bernard, A.~Lalande, C.~Zotti, F.~Cervenansky, X.~Yang, P.-A. Heng,
  I.~Cetin, K.~Lekadir, O.~Camara, M.~A.~G. Ballester \emph{et~al.}, ``Deep
  learning techniques for automatic mri cardiac multi-structures segmentation
  and diagnosis: is the problem solved?'' \emph{IEEE transactions on medical
  imaging}, vol.~37, no.~11, pp. 2514--2525, 2018.

\bibitem{gerber2017multiscale}
N.~Gerber, M.~Reyes, L.~Barazzetti, H.~M. Kjer, S.~Vera, M.~Stauber,
  P.~Mistrik, M.~Ceresa, N.~Mangado, W.~Wimmer \emph{et~al.}, ``A multiscale
  imaging and modelling dataset of the human inner ear,'' \emph{Scientific
  data}, vol.~4, p. 170132, 2017.

\bibitem{jafari2019echocardiography}
M.~H. Jafari, Z.~Liao, H.~Girgis, M.~Pesteie, R.~Rohling, K.~Gin, T.~Tsang, and
  P.~Abolmaesumi, ``Echocardiography segmentation by quality translation using
  anatomically constrained cyclegan,'' in \emph{International Conference on
  Medical Image Computing and Computer-Assisted Intervention}.\hskip 1em plus
  0.5em minus 0.4em\relax Springer, 2019, pp. 655--663.

\bibitem{cciccek20163d}
{\"O}.~{\c{C}}i{\c{c}}ek, A.~Abdulkadir, S.~S. Lienkamp, T.~Brox, and
  O.~Ronneberger, ``3d u-net: learning dense volumetric segmentation from
  sparse annotation,'' in \emph{International conference on medical image
  computing and computer-assisted intervention}.\hskip 1em plus 0.5em minus
  0.4em\relax Springer, 2016, pp. 424--432.

\bibitem{ghose2012survey}
S.~Ghose, A.~Oliver, R.~Mart{\'\i}, X.~Llad{\'o}, J.~C. Vilanova, J.~Freixenet,
  J.~Mitra, D.~Sidib{\'e}, and F.~Meriaudeau, ``A survey of prostate
  segmentation methodologies in ultrasound, magnetic resonance and computed
  tomography images,'' \emph{Computer methods and programs in biomedicine},
  vol. 108, no.~1, pp. 262--287, 2012.

\bibitem{girum2020fast}
K.~B. Girum, G.~Cr{\'e}hange, R.~Hussain, and A.~Lalande, ``Fast interactive
  medical image segmentation with weakly supervised deep learning method,''
  \emph{International Journal of Computer Assisted Radiology and Surgery},
  vol.~15, no.~9, pp. 1437--1444, 2020.

\bibitem{litjens2017survey}
G.~Litjens, T.~Kooi, B.~E. Bejnordi, A.~A.~A. Setio, F.~Ciompi, M.~Ghafoorian,
  J.~A. Van Der~Laak, B.~Van~Ginneken, and C.~I. S{\'a}nchez, ``A survey on
  deep learning in medical image analysis,'' \emph{Medical image analysis},
  vol.~42, pp. 60--88, 2017.

\bibitem{krizhevsky2012imagenet}
A.~Krizhevsky, I.~Sutskever, and G.~E. Hinton, ``Imagenet classification with
  deep convolutional neural networks,'' in \emph{Advances in neural information
  processing systems}, 2012, pp. 1097--1105.

\bibitem{he2017mask}
K.~He, G.~Gkioxari, P.~Doll{\'a}r, and R.~Girshick, ``Mask r-cnn,'' in
  \emph{Proceedings of the IEEE international conference on computer vision},
  2017, pp. 2961--2969.

\bibitem{ronneberger2015u}
O.~Ronneberger, P.~Fischer, and T.~Brox, ``U-net: Convolutional networks for
  biomedical image segmentation,'' in \emph{International Conference on Medical
  image computing and computer-assisted intervention}.\hskip 1em plus 0.5em
  minus 0.4em\relax Springer, 2015, pp. 234--241.

\bibitem{long2015fully}
J.~Long, E.~Shelhamer, and T.~Darrell, ``Fully convolutional networks for
  semantic segmentation,'' in \emph{Proceedings of the IEEE conference on
  computer vision and pattern recognition}, 2015, pp. 3431--3440.

\bibitem{xu2019deepatlas}
Z.~Xu and M.~Niethammer, ``Deepatlas: Joint semi-supervised learning of image
  registration and segmentation,'' in \emph{International Conference on Medical
  Image Computing and Computer-Assisted Intervention}.\hskip 1em plus 0.5em
  minus 0.4em\relax Springer, 2019, pp. 420--429.

\bibitem{li2019hybrid}
B.~Li, W.~J. Niessen, S.~Klein, M.~de~Groot, M.~A. Ikram, M.~W. Vernooij, and
  E.~E. Bron, ``A hybrid deep learning framework for integrated segmentation
  and registration: Evaluation on longitudinal white matter tract changes,'' in
  \emph{International Conference on Medical Image Computing and
  Computer-Assisted Intervention}.\hskip 1em plus 0.5em minus 0.4em\relax
  Springer, 2019, pp. 645--653.

\bibitem{goodfellow2014generative}
I.~Goodfellow, J.~Pouget-Abadie, M.~Mirza, B.~Xu, D.~Warde-Farley, S.~Ozair,
  A.~Courville, and Y.~Bengio, ``Generative adversarial nets,'' in
  \emph{Advances in neural information processing systems}, 2014, pp.
  2672--2680.

\bibitem{simonyan2014very}
K.~Simonyan and A.~Zisserman, ``Very deep convolutional networks for
  large-scale image recognition,'' \emph{arXiv preprint arXiv:1409.1556}, 2014.

\bibitem{badrinarayanan2017segnet}
V.~Badrinarayanan, A.~Kendall, and R.~Cipolla, ``Segnet: A deep convolutional
  encoder-decoder architecture for image segmentation,'' \emph{IEEE
  transactions on pattern analysis and machine intelligence}, vol.~39, no.~12,
  pp. 2481--2495, 2017.

\bibitem{noh2015learning}
H.~Noh, S.~Hong, and B.~Han, ``Learning deconvolution network for semantic
  segmentation,'' in \emph{Proceedings of the IEEE international conference on
  computer vision}, 2015, pp. 1520--1528.

\bibitem{szegedy2015going}
C.~Szegedy, W.~Liu, Y.~Jia, P.~Sermanet, S.~Reed, D.~Anguelov, D.~Erhan,
  V.~Vanhoucke, and A.~Rabinovich, ``Going deeper with convolutions,'' in
  \emph{Proceedings of the IEEE conference on computer vision and pattern
  recognition}, 2015, pp. 1--9.

\bibitem{he2016deep}
K.~He, X.~Zhang, S.~Ren, and J.~Sun, ``Deep residual learning for image
  recognition,'' in \emph{Proceedings of the IEEE conference on computer vision
  and pattern recognition}, 2016, pp. 770--778.

\bibitem{szegedy2017inception}
C.~Szegedy, S.~Ioffe, V.~Vanhoucke, and A.~Alemi, ``Inception-v4,
  inception-resnet and the impact of residual connections on learning,'' in
  \emph{Proceedings of the AAAI Conference on Artificial Intelligence},
  vol.~31, no.~1, 2017.

\bibitem{tu2009auto}
Z.~Tu and X.~Bai, ``Auto-context and its application to high-level vision tasks
  and 3d brain image segmentation,'' \emph{IEEE transactions on pattern
  analysis and machine intelligence}, vol.~32, no.~10, pp. 1744--1757, 2009.

\bibitem{oktay2017anatomically}
O.~Oktay, E.~Ferrante, K.~Kamnitsas, M.~Heinrich, W.~Bai, J.~Caballero, S.~A.
  Cook, A.~De~Marvao, T.~Dawes, D.~P. O‘Regan \emph{et~al.}, ``Anatomically
  constrained neural networks (acnns): application to cardiac image enhancement
  and segmentation,'' \emph{IEEE transactions on medical imaging}, vol.~37,
  no.~2, pp. 384--395, 2017.

\bibitem{girum2019deep}
K.~B. Girum, G.~Cr{\'e}hange, R.~Hussain, P.~M. Walker, and A.~Lalande, ``Deep
  generative model-driven multimodal prostate segmentation in radiotherapy,''
  in \emph{Workshop on Artificial Intelligence in Radiation Therapy}.\hskip 1em
  plus 0.5em minus 0.4em\relax Springer, 2019, pp. 119--127.

\bibitem{zotti2018convolutional}
C.~Zotti, Z.~Luo, A.~Lalande, and P.-M. Jodoin, ``Convolutional neural network
  with shape prior applied to cardiac mri segmentation,'' \emph{IEEE journal of
  biomedical and health informatics}, vol.~23, no.~3, pp. 1119--1128, 2018.

\bibitem{raghu2019transfusion}
M.~Raghu, C.~Zhang, J.~Kleinberg, and S.~Bengio, ``Transfusion: Understanding
  transfer learning for medical imaging,'' in \emph{Advances in neural
  information processing systems}, 2019, pp. 3347--3357.

\bibitem{gu2019net}
Z.~Gu, J.~Cheng, H.~Fu, K.~Zhou, H.~Hao, Y.~Zhao, T.~Zhang, S.~Gao, and J.~Liu,
  ``Ce-net: Context encoder network for 2d medical image segmentation,''
  \emph{IEEE transactions on medical imaging}, vol.~38, no.~10, pp. 2281--2292,
  2019.

\bibitem{zeng2019liver}
Q.~Zeng, D.~Karimi, E.~H. Pang, S.~Mohammed, C.~Schneider, M.~Honarvar, and
  S.~E. Salcudean, ``Liver segmentation in magnetic resonance imaging via mean
  shape fitting with fully convolutional neural networks,'' in
  \emph{International Conference on Medical Image Computing and
  Computer-Assisted Intervention}.\hskip 1em plus 0.5em minus 0.4em\relax
  Springer, 2019, pp. 246--254.

\bibitem{girum2020deep}
K.~B. Girum, A.~Lalande, R.~Hussain, and G.~Crehange, ``A deep learning method
  for real-time intraoperative us image segmentation in prostate
  brachytherapy,'' \emph{International Journal of Computer Assisted Radiology
  and Surgery}, vol.~15, no.~9, pp. 1467--1476, 2020.

\bibitem{chen2019learning}
C.~Chen, C.~Biffi, G.~Tarroni, S.~Petersen, W.~Bai, and D.~Rueckert, ``Learning
  shape priors for robust cardiac mr segmentation from multi-view images,'' in
  \emph{International Conference on Medical Image Computing and
  Computer-Assisted Intervention}.\hskip 1em plus 0.5em minus 0.4em\relax
  Springer, 2019, pp. 523--531.

\bibitem{ravishankar2017learning}
H.~Ravishankar, R.~Venkataramani, S.~Thiruvenkadam, P.~Sudhakar, and V.~Vaidya,
  ``Learning and incorporating shape models for semantic segmentation,'' in
  \emph{International conference on medical image computing and
  computer-assisted intervention}.\hskip 1em plus 0.5em minus 0.4em\relax
  Springer, 2017, pp. 203--211.

\bibitem{larrazabal2020post}
A.~J. Larrazabal, C.~Mart{\'\i}nez, B.~Glocker, and E.~Ferrante, ``Post-dae:
  Anatomically plausible segmentation via post-processing with denoising
  autoencoders,'' \emph{IEEE Transactions on Medical Imaging}, vol.~39, no.~12,
  pp. 3813--3820, 2020.

\bibitem{painchaud2020cardiac}
N.~Painchaud, Y.~Skandarani, T.~Judge, O.~Bernard, A.~Lalande, and P.-M.
  Jodoin, ``Cardiac segmentation with strong anatomical guarantees,''
  \emph{IEEE Transactions on Medical Imaging}, vol.~39, no.~11, pp. 3703--3713,
  2020.

\bibitem{anderson2018bottom}
P.~Anderson, X.~He, C.~Buehler, D.~Teney, M.~Johnson, S.~Gould, and L.~Zhang,
  ``Bottom-up and top-down attention for image captioning and visual question
  answering,'' in \emph{Proceedings of the IEEE conference on computer vision
  and pattern recognition}, 2018, pp. 6077--6086.

\bibitem{schuster1997bidirectional}
M.~Schuster and K.~K. Paliwal, ``Bidirectional recurrent neural networks,''
  \emph{IEEE transactions on Signal Processing}, vol.~45, no.~11, pp.
  2673--2681, 1997.

\bibitem{schlemper2019attention}
J.~Schlemper, O.~Oktay, M.~Schaap, M.~Heinrich, B.~Kainz, B.~Glocker, and
  D.~Rueckert, ``Attention gated networks: Learning to leverage salient regions
  in medical images,'' \emph{Medical image analysis}, vol.~53, pp. 197--207,
  2019.

\bibitem{sinha2020multi}
A.~Sinha and J.~Dolz, ``Multi-scale self-guided attention for medical image
  segmentation,'' \emph{IEEE Journal of Biomedical and Health Informatics},
  vol.~25, no.~1, pp. 121--130, 2020.

\bibitem{li2018referring}
R.~Li, K.~Li, Y.-C. Kuo, M.~Shu, X.~Qi, X.~Shen, and J.~Jia, ``Referring image
  segmentation via recurrent refinement networks,'' in \emph{Proceedings of the
  IEEE Conference on Computer Vision and Pattern Recognition}, 2018, pp.
  5745--5753.

\bibitem{chen2016combining}
J.~Chen, L.~Yang, Y.~Zhang, M.~Alber, and D.~Z. Chen, ``Combining fully
  convolutional and recurrent neural networks for 3d biomedical image
  segmentation,'' in \emph{Advances in neural information processing systems},
  2016, pp. 3036--3044.

\bibitem{alom2019recurrent}
M.~Z. Alom, C.~Yakopcic, M.~Hasan, T.~M. Taha, and V.~K. Asari, ``Recurrent
  residual u-net for medical image segmentation,'' \emph{Journal of Medical
  Imaging}, vol.~6, no.~1, p. 014006, 2019.

\bibitem{wang2019recurrent}
W.~Wang, K.~Yu, J.~Hugonot, P.~Fua, and M.~Salzmann, ``Recurrent u-net for
  resource-constrained segmentation,'' in \emph{Proceedings of the IEEE
  International Conference on Computer Vision}, 2019, pp. 2142--2151.

\bibitem{lin2017refinenet}
G.~Lin, A.~Milan, C.~Shen, and I.~Reid, ``Refinenet: Multi-path refinement
  networks for high-resolution semantic segmentation,'' in \emph{Proceedings of
  the IEEE conference on computer vision and pattern recognition}, 2017, pp.
  1925--1934.

\bibitem{wangfrnet}
D.~Wang, G.~Hu, and C.~Lyu, ``Frnet: an end-to-end feature refinement neural
  network for medical image segmentation,'' \emph{The Visual Computer}, pp.
  1--12, 2020.

\bibitem{geirhos2018imagenet}
R.~Geirhos, P.~Rubisch, C.~Michaelis, M.~Bethge, F.~A. Wichmann, and
  W.~Brendel, ``Imagenet-trained cnns are biased towards texture; increasing
  shape bias improves accuracy and robustness,'' in \emph{International
  Conference on Learning Representations}, 2018.

\bibitem{shama2019adversarial}
F.~Shama, R.~Mechrez, A.~Shoshan, and L.~Zelnik-Manor, ``Adversarial feedback
  loop,'' in \emph{Proceedings of the IEEE International Conference on Computer
  Vision}, 2019, pp. 3205--3214.

\bibitem{huh2019feedback}
M.~Huh, S.-H. Sun, and N.~Zhang, ``Feedback adversarial learning: Spatial
  feedback for improving generative adversarial networks,'' in
  \emph{Proceedings of the IEEE Conference on Computer Vision and Pattern
  Recognition}, 2019, pp. 1476--1485.

\bibitem{hu2018squeeze}
J.~Hu, L.~Shen, and G.~Sun, ``Squeeze-and-excitation networks,'' in
  \emph{Proceedings of the IEEE conference on computer vision and pattern
  recognition}, 2018, pp. 7132--7141.

\bibitem{kingma2014adam}
D.~P. Kingma and J.~Ba, ``Adam: A method for stochastic optimization,''
  \emph{arXiv preprint arXiv:1412.6980}, 2014.

\bibitem{leclerc2019deep}
S.~Leclerc, E.~Smistad, J.~Pedrosa, A.~{\O}stvik, F.~Cervenansky, F.~Espinosa,
  T.~Espeland, E.~A.~R. Berg, P.-M. Jodoin, T.~Grenier \emph{et~al.}, ``Deep
  learning for segmentation using an open large-scale dataset in 2d
  echocardiography,'' \emph{IEEE transactions on medical imaging}, vol.~38,
  no.~9, pp. 2198--2210, 2019.

\bibitem{whitley2002statistics}
E.~Whitley and J.~Ball, ``Statistics review 6: Nonparametric methods,''
  \emph{Critical care}, vol.~6, no.~6, p. 509, 2002.

\bibitem{hussain2020augmented}
R.~Hussain, A.~Lalande, K.~B. Girum, C.~Guigou, and A.~B. Grayeli, ``Augmented
  reality for inner ear procedures: visualization of the cochlear central axis
  in microscopic videos,'' \emph{International Journal of Computer Assisted
  Radiology and Surgery}, vol.~15, no.~10, pp. 1703--1711, 2020.

\end{thebibliography}

\end{document}